\def\be{\begin{equation}}
\def\ee{\end{equation}}
\def\bea{\begin{eqnarray}}
\def\eea{\end{eqnarray}}

\documentclass{ws-p9-75x6-50}

\begin{document}

\title{Transport Anomalies and Marginal Fermi-Liquid Effects at a Quantum
       Critical Point}
\author{D. Belitz and Sharon L. Sessions}
\address{Department of Physics and Materials Science Institute, University 
         of Oregon, Eugene, OR 97403, USA 
         (e-mail: belitz@greatwhite.uoregon.edu)}

\author{T.R. Kirkpatrick}
\address{Insitute for Physical Science and Technology, and Department of 
         Physics, University of Maryland, College Park, MD 20742}

\author{M.T. Mercaldo}
\address{Dipartimento di Fisica ``E.R. Caianello'' and Istituto Nazionale 
         di Fisica per la Materia, Universit{\'a} di Salerno, I-84081 
         Baronissi (SA), Italy}

\author{R. Narayanan and Thomas Vojta}
\address{Theoretical Physics, University of Oxford, OX1 3NP, UK}

\maketitle

\abstracts{
The behavior of the conductivity and the density of states, as well as the
phase relaxation time, of disordered itinerant electrons across a quantum
ferromagnetic transition is discussed. It is shown that critical fluctuations
lead to anomalies in the temperature and energy dependence of the conductivity
and the tunneling density of states, respectively, 
that are stronger than the usual
weak-localization anomalies in a disordered Fermi liquid. This can be used
as an experimental probe of the quantum critical behavior. The energy
dependence of the phase relaxation time at criticality is shown to be
that of a marginal Fermi liquid.
}

\section{Introduction}
\label{sec:1}
\subsection{Disordered Fermi Liquids}
\label{subsec:1.1}

It is well known that the interplay between quenched disorder and
electron-electron interaction effects lead to non-analytic dependencies
of various observables on control parameters. For instance, in three-dimensions
(which we will assume throughout) 
the conductivity has a square-root temperature dependence,\cite{AAL}
\begin{equation}
\sigma(T) = \sigma_0\,\left[1 + {\rm const.}\times\sqrt{T}\right]\quad,
\label{eq:1}
\end{equation}
the density of states has a square-root energy dependence,\cite{AA}
\begin{equation}
N(\epsilon) = N_F\,\left[1 + {\rm const.}\times\sqrt{\epsilon}\,\right]\quad,
\label{eq:2}
\end{equation}
and the phase relaxation rate has an $\epsilon^{3/2}$ energy 
dependence,\cite{Schmid}
\begin{equation}
\tau_{\rm ph}^{-1}(\epsilon) = {\rm const.}\times\epsilon^{3/2}\quad.
\label{eq:3}
\end{equation}
These effects, and similar other ones, are collectively known
as `weak-localization effects'. They were originally derived by means of
perturbation theory,\cite{WL_reviews}
and were later understood to be the leading corrections to scaling at a stable
renormalization-group fixed point that describes a disordered Fermi 
liquid\cite{us_fermions}. They can also be interpreted as an example of
`generic scale invariance', i.e. the occurrence of power-law correlations
in an entire phase, rather than only at an isolated critical 
point.\cite{us_ernst} 

\subsection{Quantum Critical Behavior of Itinerant Ferromagnets}
\label{subsec:1.2}

Usually, a ferromagnetic phase transition is triggered by changing the
temperature through the Curie point. The resulting critical behavior is
caused by thermal fluctuations that become large near the critical point,
and is referred to as `thermal' or `classical' critical behavior. However,
a system with a low Curie temperature can also be driven into the
paramagnetic phase by means of a non-thermal control parameter, such as
pressure, or composition, at zero or very low temperature. An example is
MnSi, which undergoes a ferromagnet-to-paramagnet transition at $T=0$
and a hydrostatic pressure $p_c\approx 15$ kbar.\cite{Pfleiderer_et_al}
This is an example of a `quantum critical point'.\cite{Hertz,Sachdev} 
The critical
behavior is driven by quantum fluctuations, and it is therefore different
from that at the corresponding thermal phase transition.

The critical behavior at the quantum ferromagnetic transition of disordered
itinerant electrons has recently
been determined exactly,\cite{us_fm_local} and was found to consist of
power laws\cite{us_fm} with log-log-normal corrections to scaling.
For instance, the magnetization behaves like
\begin{equation}
m(t) \propto t^2\,e^{(\ln\ln(1/t))^2}\quad,
\label{eq:4}
\end{equation}
where $t$ is the dimensionless distance from the critical point at $T=0$,
e.g., $t=\vert p-p_c\vert/p_c$. In writing Eq.\ (\ref{eq:4}), we have omitted
constant prefactors in the exponent.

Equation (\ref{eq:4}) determines the critical exponent $\beta$, which
is defined by the behavior of the magnetization as a function of $t$,
$m(t)\propto t^{\beta}$. A convenient way to account for the fact that,
due to the logarithmic corrections to scaling, the asymptotic critical
behavior is not simply given by power laws, is to formally make the
critical exponents scale dependent. We thus write
\begin{equation}
\beta = 2 + \ln g(\ln b)/\ln b\quad,
\label{eq:5}
\end{equation}
with $b$ an arbitrary renormalization group length scale factor, and
$g$ a function that, for large arguments and omitting constant prefactors,
can be represented by $g(x) \approx \exp[(\ln x)^2]$. More generally, one needs
three independent critical exponents to describe quantum critical 
behavior. One of these is the dynamical critical exponent $z$, which
determines the scaling behavior of frequency and temperature. It 
reads\cite{us_fm_local}
\begin{equation}
z = 3 + \ln g(\ln b)/\ln b\quad.
\label{eq:6}
\end{equation}
In addition
to Eq.\ (\ref{eq:6}) there is a second time scale in the problem which plays
an important role in the complete theory\cite{us_fm_local} but is not
crucial for our present purposes. 

For the second static exponent we choose 
the correlation length exponent $\nu$,\cite{us_fm_local}
\begin{equation}
1/\nu = 1 + \ln g(\ln b)/\ln b\quad.
\label{eq:7}
\end{equation}
Finally, for later reference, we note that the scale dimension of the leading
irrelevant operator in a disordered electron system, which we denote by $u$, 
is given by $-1$.\cite{us_fermions}

Notice that the critical behavior is not mean-field
like, in contrast to the prediction by Hertz.\cite{Hertz}
The physical reason for this breakdown of Hertz's order parameter field
theory lies in the existence of soft modes in itinerant ferromagnets, 
viz. particle-hole excitations, that are distinct from the order parameter
fluctuations, but couple sufficiently strongly to the latter to influence
the critical behavior.

\section{Feedback of critical behavior on weak-localization effects}
\label{sec:2}

The particle-hole excitations that are responsible for the non-mean field
critical behavior described above are the same soft modes that
lead to the weak-localization 
nonanalyticities. One should therefore expect the critical behavior to
change the latter via feedback effects. This is indeed the case, as has
been shown in Ref.\ \citelow{us_Letter}.

The easiest way to determine this effect is by means of scaling arguments.
Let us start with the electrical conductivity. This transport coefficient
is unrelated to magnetism, so at a magnetic phase transition one expects
its scale dimension to be zero. We thus have a scaling or homogeneity law
\begin{equation}
\sigma(t,T) = F_{\sigma}(t\,b^{1/\nu},T\,b^z,u\,b^{-1})\quad,
\label{eq:8}
\end{equation}
with $F_{\sigma}$ a scaling function. Using the fact
that $F_{\sigma}(0,1,x)$ is an analytic function of $x$,\cite{us_fermions} 
and putting $b=T^{-1/3}$, we obtain at criticality, $t=0$,
\begin{equation}
\sigma(t=0,T) = \sigma_0\left[1 + c_{\sigma}\left[(T/T_F)\,
   g(\ln(T_F/T))\right]^{1/3} + O(\sqrt{T})\right]\quad.
\label{eq:9}
\end{equation}
Here $\sigma_0$ is the disordered Fermi-liquid value of $\sigma$, $T_F$ is
the Fermi temperature, and $c_{\sigma}$ is a disorder dependent constant.

Similarly, the leading correction to the density of states is given by an
integral over a four-fermion correlation function\cite{AA,us_fermions} whose
diffusive dynamics lead to $\Delta N\sim b^{-1}$. We
thus have
\begin{equation}
\Delta N(t,\epsilon) = b^{-1}\,F_N(t\,b^{-1/\nu},\epsilon\,b^z)\quad,
\label{eq:10}
\end{equation}
with $F_N$ another scaling function. At criticality this yields
\begin{equation}
N(t=0,\epsilon) = N_F\,\left[1 + c_N\,\left[(\epsilon/\epsilon_F)\,
   g(\ln(\epsilon_F/\epsilon))\right]^{1/3} + O(\sqrt{\epsilon})\right]\
   \quad,
\label{eq:11}
\end{equation}
with $N_F$ the disordered Fermi liquid value of the density of states at the
Fermi level, and $\epsilon_F$ the Fermi energy.

Finally, in a disordered electron system the phase relaxation rate 
scales like a wavenumber squared, and thus has a scale dimension of $2$.
This implies
\begin{equation}
\tau_{\rm ph}^{-1}(t,\epsilon) = b^{-2}\,F_{\rm ph}(t\,b^{-1/\nu},
   \epsilon\,b^z, u\,b^{-1})\quad,
\label{eq:12}
\end{equation}
with $F_{\rm ph}$ a third scaling function. The leading irrelevant variable
$u$ represents interaction effects that are necessary for any dephasing, and
the rate is therefore proportional to $u$. At criticality, and with a
constant $c_{\rm ph}$, this leads to
\begin{equation}
\tau_{\rm ph}^{-1}(t=0,\epsilon) = c_{\rm ph}\,\epsilon\,g\left(
   \ln(\epsilon_F/\epsilon)\right) + O(\epsilon^{3/2})\quad.
\label{eq:13}
\end{equation}

Equations (\ref{eq:11}) and (\ref{eq:13}) remain valid if $\epsilon$ is
replaced by $T$ (at $\epsilon = 0$), and Eq.\ (\ref{eq:9}) remains valid
if $T$ is replaced by the frequency $\Omega$.

\section{Discussion}
\label{sec:3}

As we have made plausible above, and have shown in more detail in
Refs.\ \citelow{us_Letter,us_fm_local}, the transport and thermodynamic
properties of disordered electrons show remarkable anomalies at a quantum
ferromagnetic critical point. The conductivity and the density of states
show nonanalytic dependencies on the temperature or frequency/energy
that are similar to, but stronger than, the usual weak-localization anomalies.
The crossover from the square-root dependence of the latter to the cube-root
of the former should be easily observable experimentally. Such an experiment
would provide a check of the theoretically predicted value $z=3$ of the
dynamical critical exponent. For both quantities, there are strong
multiplicative corrections to scaling, which in an experiment with a limited
dynamical range would result in an effective exponent that is different
from $3$. For the phase relaxation time, we have found an even more drastic
effect: Schmid's $\epsilon^{3/2}$ (or $T^{3/2}$) law gets replaced by a
linear energy (or temperature) dependence with a multiplicative logarithmic
correction. The electron system at the quantum ferromagnetic critical point
thus has single-particle properties that are those of a marginal Fermi
liquid.\cite{MFL}

\section*{Acknowledgments}
Part of this work was performed at the Aspen Center for Physics, and
supported by the NSF under grant Nos. DMR-98-70597 and DMR-99-75259,
by the DFG under grant No. Vo659/3, and by the EPSRC under grant No.
GR/M 04426.

\end{document}